\documentclass[reprint,aps,prl]{revtex4-2}
\usepackage{amsmath}
\allowdisplaybreaks[4]
\usepackage{multirow}
\usepackage{amsfonts}
\usepackage{xcolor}
\usepackage{comment}
\usepackage{graphicx}
\usepackage{amsmath}
\usepackage{siunitx}
\usepackage{lineno}
\usepackage{float}
\usepackage{ulem}
\usepackage[
    pdfauthor={Teng Zhang},
    pdftitle={A two-carrier scheme: evading 3dB quantum penalty of heterodyne readout in gravitational-wave detectors},
    colorlinks=true,
    linkcolor=black,
    urlcolor=blue,
    citecolor=blue
]{hyperref}

\DeclareSIUnit{\sqrthz}{\ensuremath{\sqrt{\text{\hertz}}}}

\begin{document}

\renewcommand\texteuro{FIXME}

\allowdisplaybreaks[4]
\title{A two-carrier scheme: evading the 3dB quantum penalty of heterodyne readout in gravitational-wave detectors }

\author{Teng Zhang$^1$}
\author{Philip Jones$^1$}
\author{Ji\v{r}\'{i} Smetana}
\author{Haixing Miao$^1$}
\author{Denis Martynov$^1$}
\author{Andreas Freise$^{1,2,3}$}
\author{Stefan W. Ballmer$^4$}

\affiliation{$^1$School of Physics and Astronomy, and Institute of Gravitational Wave Astronomy, University of Birmingham, Edgbaston, Birmingham B15\,2TT, United Kingdom}
\affiliation{$^2$Department of Physics and Astronomy, VU Amsterdam, De Boelelaan 1081, 1081, HV, Amsterdam, The Netherlands}
\affiliation{$^3$Nikhef, Science Park 105, 1098, XG Amsterdam, The Netherlands}
\affiliation{$^4$Syracuse University, Syracuse, NY 13244, USA}
\begin{abstract}
Precision measurements using traditional heterodyne readout suffer a 3dB quantum noise penalty compared with homodyne readout. 
The extra noise is caused by the quantum fluctuations in the image vacuum.
We propose a two-carrier gravitational-wave detector design that evades the 3dB quantum penalty of heterodyne readout.
We further propose a new way of realising frequency-dependent squeezing utilising two-mode squeezing in our scheme. It naturally achieves more precise audio frequency signal measurements with radio frequency squeezing. In addition, the detector is compatible with other quantum nondemolition techniques. 
\end{abstract}
\maketitle
\textit{Introduction ---}
Since 2015, laser interferometric gravitational-wave detectors have made a series of direct observations of gravitational waves from mergers of binary black holes and neutron stars\,\cite{GW150914,GW170817,PhysRevX.9.031040}. They have opened a new observational window into the universe and provided significant inputs to many scientific fields. 
In the detector, to translate the electromagnetic sidebands into a measurable electrical signal, a readout scheme is required, which is also fundamental for determining the sensitivity of the detector\,\cite{Fritschel2014}. 
\begin{figure}[b]
\centering
  \includegraphics[width=0.9\columnwidth]{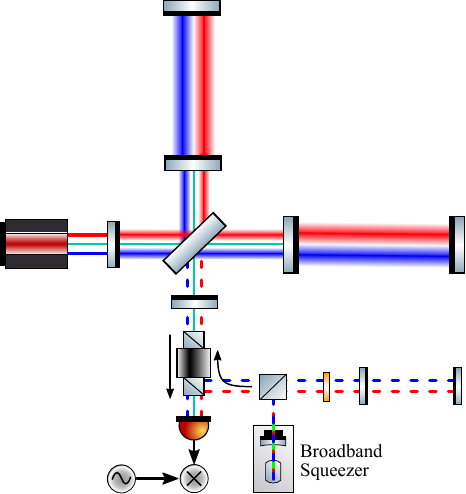}
\caption{Schematic of the two-carrier gravitational-wave detector with heterodyne readout. The red, blue and cyan lasers correspond to the carriers at $\omega_1$, $\omega_2$ and the LO at $\omega_L=(\omega_1+\omega_2)/2$ respectively. The three beams are spatially overlapped in the detector. The broadband squeezer is pumped at $2\omega_L$ and has a bandwidth from $\omega_1-\Omega$ to $\omega_2+\Omega$.}
\label{fig:detector}
\end{figure}

Heterodyne readout is widely implemented in precision measurements, \textit{e.g.}, for the stabilisation of laser frequencies and optical cavities (also known as the Pound-Drever-Hall technique\,\cite{DreverPDH, Black_PDH_2001}) and for quantum squeezing characterisation due to its natural immunity to the low-frequency laser noise\,\cite{Gea-Banacloche1987,PhysRevA.57.3898,Marino:07,Li:15,Feng:16}.
Compared with homodyne readout, heterodyne readout suffers 3dB noise penalty as the scheme picks up the vacuum fields above and below the local oscillator (LO). The additional field that does not coincide with the signal is called the image vacuum\, \cite{6771861,Yuen:83,QuantumHomHet,RevModPhys.66.481}. The noise penalty is a direct and necessary consequence of the Heisenberg uncertainty principle when all quadratures are allowed to be measured simultaneously\,\cite{PhysRevD.67.122005}. In the first-generation of gravitational wave detectors, a heterodyne readout with two balanced radio frequency (RF) sidebands was used\,\cite{Hild_2009}, reducing the factor of 2 (3dB) quantum penalty to a factor of 1.5. In that scheme, the vacuum fields which are twice the modulation frequency away from the carrier couple to the readout channel\,\cite{PhysRevD.67.122005}.
In subsequent detector upgrades the readout scheme was switched to DC readout~\cite{Hild_2009, Fricke_2012}, a variant of homodyne readout. The LO in the DC readout is derived by slightly detuning the arm cavities, which offsets the interferometer from a perfect dark fringe. DC readout has the advantage of a straightforward implementation without needing an external LO. However, the dark-fringe offset induces extra couplings of technical noises and is not ideal for future-generation gravitational wave detectors\,\cite{PhysRevLett.120.141102}. A balanced homodyne readout can eliminate the dark fringe offset by introducing a spatially separated LO\,\cite{Fritschel2014}. This requires auxiliary optics on the LO path and additional output optical mode cleaner. Meanwhile, heterodyne readout is used in current detectors for the stabilisation of auxiliary degrees of freedom, \textit{e.g.}, the lengths of recycling cavities \cite{Arain_RECYCLING_2008,Izumi_2016}. 

\begin{figure}[t]
\centering
  \includegraphics[width=1\columnwidth]{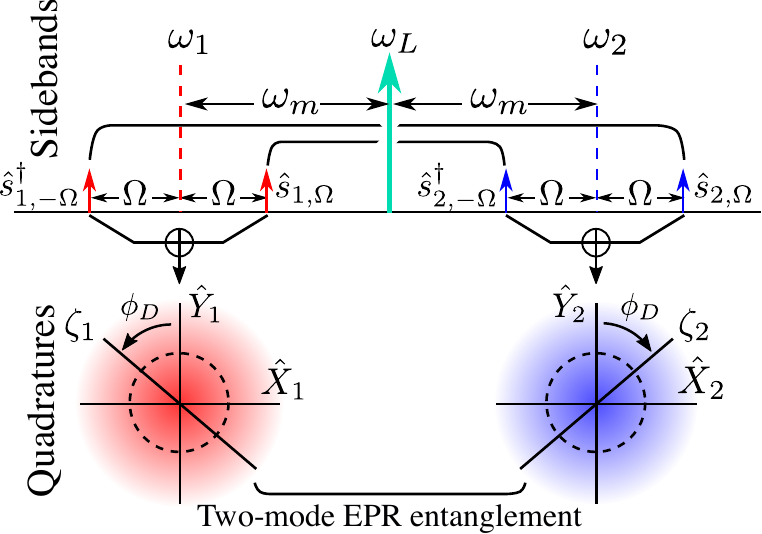}
\caption{Two-carrier heterodyne readout with two-mode squeezing: sideband picture (top) and quadrature picture (bottom). The two carriers are at frequencies $\omega_1$ and $\omega_2$. The LO field is in the middle at $\omega_L=(\omega_1+\omega_2)/2$. The squeezer is pumped at $2\omega_L$, entangling sideband pairs symmetric around the LO field at $\omega_L$. This leads to a two-mode EPR entanglement between quadratures on $\zeta_{1}$ and $\zeta_2$.  $\Omega$ is audio frequency and $\omega_m$ is the separation between the LO and each carrier.}
\label{fig:Squeezing}
\end{figure}

Is there a way to evade the fundamental quantum noise penalty with heterodyne readout? It was found that the noise penalty of heterodyne readout could be evaded if the image vacuum fields can be excited to contain coherent signal flux\,\cite{QuantumHomHet, Fan:15, PhysRevLett.113.263603}. Inspired by this finding, we give a new gravitational wave detector scheme that includes two carriers at $\omega_1$ and $\omega_2$ with a beam at $\omega_L=(\omega_1+\omega_2)/2$ serving as the heterodyne LO. The three beams are evenly separated by an RF $\omega_m$. The schematic of the design is shown in Fig.~\ref{fig:detector}. The two carriers resonate and the LO anti-resonates in the arm cavities. The LO resonates in recycling cavities. This new design with heterodyne readout will lead to same quantum-limited sensitivity as with homodyne readout and the same total arm power.

Another highlight of the two-carrier detector with heterodyne readout is the simplicity of generating quantum squeezing. Most gravitational-wave signals from compact binary system detected by ground-based detectors are within audio-band, from several hertz to several kilohertz.
However, at audio frequencies, excess noises are significant due to the parasitic interferences from scattered light\,\cite{Vahlbruch_2007,Chua_2014}. 
In general, even though the signal field is on dark fringe, bright LO is required for photon detection and can introduce audio-band scattering to the audio-band squeezer.
We will show that instead of observing audio-band squeezing \cite{NPSqueezing, AasiSqueezing} around LO frequency, radio frequency squeezing in a broadband two-mode quantum state is sufficient in our configuration,\textit{i.e.} low-frequency signals measurement with high-frequency squeezing\,\cite{Zhai:12,PhysRevA.96.023808}. Note that the audio-band noises with respect to the carriers cannot be omitted. 

\textit{Heterodyne readout and two-mode squeezing ---}
In a single sideband heterodyne readout, two vacuum fileds, $\hat{s}_{1,\pm\Omega},\hat{s}_{2,\pm\Omega}$ around frequencies $\omega_1$ and $\omega_2$ are measured as shown in Fig.~ \ref{fig:Squeezing}. 
The eventual photocurrent containing the signal and noise can be derived as
\begin{equation}\label{eq:photocurrent}
\begin{split}
I\propto \hat{s}_{1,- \Omega}^{\dagger}e^{i(\phi_L-\phi_D)}+
\hat{s}_{2,- \Omega}^{\dagger}e^{i(\phi_L+\phi_D)}\\
+\hat{s}_{1, \Omega}e^{-i(\phi_L-\phi_D)}+
\hat{s}_{2, \Omega}e^{-i(\phi_L+\phi_D)}\,,
\end{split}
\end{equation}
where $\phi_L$ is the LO phase and is assumed to be ${\pi}/{2}$ in this work, $\phi_D$ is the demodulation phase.
In the two-photon formalism \cite{PhysRevA.31.3068, PhysRevA.31.3093}, the photocurrent is proportional to the combined quadrature\,\cite{zhang2020quantum}
\begin{equation}\label{eq:Quadrature}
\hat Q_{\zeta}=\mathbf{H}_{\zeta}\cdot\begin{bmatrix}
\hat{X}_{1}&\hat{Y}_{1}&\hat{X}_{2}&\hat{Y}_{2}
\end{bmatrix}^{\rm T}\,,
\end{equation}
with $\mathbf{H}_{\zeta}\equiv [\cos \zeta_1,\sin \zeta_1,\cos \zeta_2,\sin \zeta_2]$\,.
Here $\hat{X}_{j}, \hat{Y}_{j}$ ($j=1,2$) represent the amplitude and phase quadrature of the sidebands $\hat{s}_{1,\pm\Omega}$ and $\hat{s}_{2,\pm\Omega}$, respectively.
$\zeta_{j}$ defines the measurement quadrature, 
\begin{equation}\label{eq:zeta}
\zeta_{1}=\phi_L-\phi_D\,,\zeta_{2}=\phi_L+\phi_D\,.
\end{equation}
We normalise the shot noise spectral density for the vacuum state to be 1. With the pumping laser frequency of the squeezer at $2\omega_L$, the quadratures of two modes at $\omega_1$ and $\omega_2$ are correlated\,\cite{Gea-Banacloche1987, PhysRevA.67.054302,SCHNABEL20171, zhang2020quantum}, of which the correlation is quantified by the covariance matrix,
\begin{equation}\label{eq:Vepr}
\mathbb{V} = \begin{bmatrix}
\alpha &0 & \beta & 0\\
0 &\alpha &0& -\beta \\
\beta&0 &\alpha & 0\\
0& -\beta &0 &\alpha
\end{bmatrix}\,,
\end{equation}
where $\alpha=\cosh 2r ,\beta=\sinh 2r$ and $r$ is the phase squeezing factor. The spectral density of the combined quadrature $\hat Q_{\zeta}$ in Eq.~\eqref{eq:Quadrature} is given by
\begin{equation}
\mathbf{H}_{\zeta}\mathbb{V}\mathbf{H}_{\zeta}^{\rm T}=2\alpha-2\beta=2e^{-2r}\,,
\end{equation}
which is a natural result of the EPR entanglement between quadratures of the two modes\,\cite{Marino:07, Ma:2017aa, Danilishin2019,EPRJan}. This is depicted clearly in Fig.~\ref{fig:Squeezing}. This two-mode quantum state can be achieved with a broadband squeezer, which in principle gives constant squeezing within the bandwidth from $\omega_1-\Omega$ to $\omega_2+\Omega$. However, in our scheme, squeezing only around $\omega_1$ and $\omega_2$---\textit{i.e.}\ $\omega_m$ away from half of the squeezer pumping frequency---is required to be observed. 
This means that although we need a broadband squeezer, good squeezing around $\omega_L$ is not required to be observed.
In a single-carrier scheme with heterodyne readout, in which only one of the two modes takes signal, the noise to signal ratio is $2e^{-2r}$, where the signal part is normalised to be 1. The factor of 2 here corresponds to the well-known 3dB quantum penalty. In the two carrier scheme, the same total power is divided into two carriers equally, and the noise to signal ratio becomes $2e^{-2r}/(\sqrt{2}/2+\sqrt{2}/2)^2=e^{-2r}$, which demonstrates the evasion of the 3dB penalty. 

\begin{figure}[t]
\centering
  \includegraphics[width=1\columnwidth]{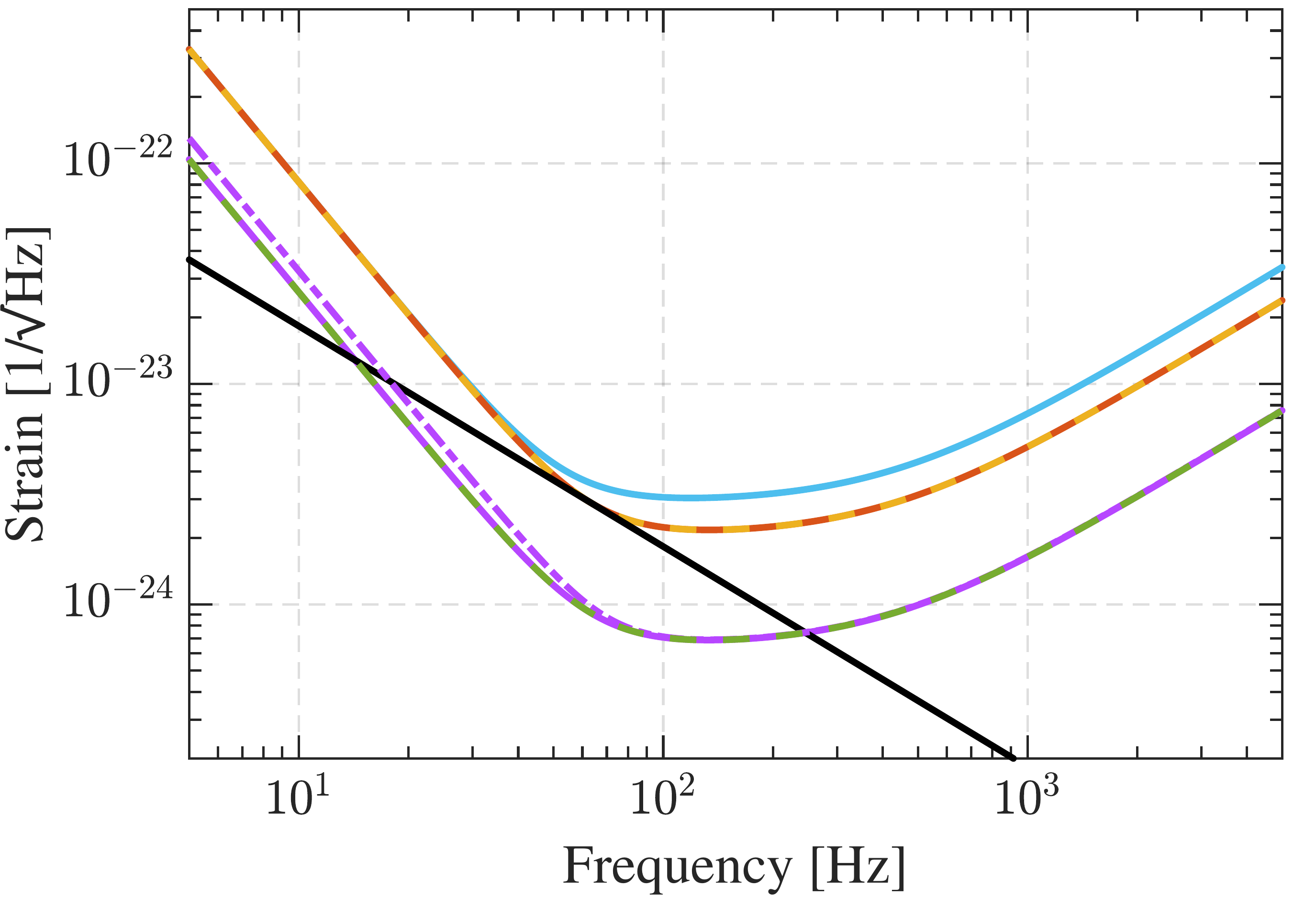}
\caption{Quantum-limited sensitivity of different configurations. The blue curve corresponds to the case of a single-carrier detector with heterodyne readout. The orange curve corresponds to the two-carrier detector, which perfectly overlaps with the dashed yellow curve for homodyne readout. The solid purple curve is the sensitivity of the two-carrier detector with 10dB squeezing, which perfectly overlaps with the dashed green curve for homodyne readout with the same squeezing. The dot-dashed purple curve corresponds to a 15\% power imbalance between the two carriers while the total power stays constant. The black curve is the Standard Quantum Limit. The detector and filter cavity parameters used in the two-carrier detector are the same as those of Advanced LIGO\,\cite{LIGO2015, PhysRevLett.124.171102} and Advanced LIGO upgrade\,\cite{T1800042}.}
\label{fig:Strain}
\end{figure}

\textit{Quantum  noise of the detector ---} So far, we have been focusing on 
shot noise. Inside the interferometer, the two sideband modes interact with the test mass through the radiation pressure force by beating with the two carriers, which also introduces radiation pressure noise. When the interferometer is tuned with equal power in two carriers, the optomechanical factors describing the interaction of the modes with the test mass mirrors and their cross correlation are identical. They are equal to half of the optomechanical factor $\cal K$
defined in Ref.\,\cite{kimble2001}. Note that here, because $\omega_m$ is much smaller than $\omega_j$, we neglect the effect of the difference in the wavelength of two carriers. The input-output transfer matrix
$\mathbb{T}$ for the quadratures of these two modes and their response vector $\mathbf{R}$ of the interferometer to the gravitational wave strain can be derived as\,\footnote{See Supplemental Material for detailed derivations, which includes Refs.\cite{PhysRevD.78.062003,PhysRevD.67.062002}}
\begin{equation}\label{eq:T}
\mathbb{T}=e^{2i\Phi}\begin{bmatrix}
1&0&0&0\\-\mathcal{K}/2&1&-\mathcal{K}/2&0\\
0&0&1&0\\
-\mathcal{K}/2&0&-\mathcal{K}/2&1
\end{bmatrix},\;\mathbf{R}=\frac{e^{i\Phi}}{h_{\rm SQL}}\begin{bmatrix}
0\\\sqrt{\mathcal{K}}\\
0\\
\sqrt{\mathcal{K}}
\end{bmatrix}\,.
\end{equation}
Here $\Phi={\rm atan}({\Omega}/{\gamma})$ is the phase of the sidebands at frequency $\Omega$ acquired by reflection from the interferometer with an effective bandwidth $\gamma$. The opto-mechanical
factor is 
\begin{equation}
\mathcal{K}=\frac{16 \omega_L P\gamma}{mcL\Omega^2(\gamma^2+\Omega^2)}\,,
\end{equation}
where $m$ is the mass of each mirror, and $P$ is the total circulating power in the arm cavity. The Standard Quantum Limit of the detector in strain is $h_{\rm SQL}=\sqrt{8\hbar/(m\Omega^2L^2)}$.

The low-frequency radiation pressure noise can be 
improved by using frequency-dependent squeezing\,\cite{kimble2001}. The filter cavity provides frequency dependent quadrature rotations $\theta_1, \theta_2$ for the two modes, which can be described by\,\cite{SD2012}
\begin{equation}
\mathbb{P}_{\theta}=\begin{bmatrix}
\cos \theta_1 &-\sin \theta_1 &0&0\\
\sin \theta_1 &\cos \theta_1 &0&0\\
0&0&\cos \theta_2 &-\sin \theta_2 \\
0&0&\sin \theta_2 &\cos \theta_2 
\end{bmatrix}\,.
\end{equation}
The quantum noise spectral density of heterodyne readout is given by
\begin{equation}
\label{eq:Stotal}
S_{hh}= \frac{ \mathbf{H}_{\zeta} \mathbb{T} \mathbb{P}_{\theta} \mathbb{V} \mathbb{P}_{\theta}^{\rm T}\mathbb{T}^{\dagger}\mathbf{H}_{\zeta}^{\rm T}}{|\mathbf{H_{\zeta}}\mathbf{R}|^2}\,.
\end{equation}
When the frequency-dependent rotation angle satisfies
\begin{equation}\label{eq:theta}
\cos(\theta_1+\theta_2)=\frac{1-\mathcal{K}^2}{1+\mathcal{K}^2}\,,
\end{equation}
the noise spectrum reaches the minimal value
\begin{equation}
S_{hh}^{\rm min}=\frac{h^2_{\rm SQL}}{2}\frac{\mathcal{K}^2+1}{\mathcal{K}}e^{-2r}\,.
\end{equation}
In a special case of Eq.~\ref{eq:theta}, $\theta_1=\theta_2={\rm atan}\, \mathcal{K}$, the required frequency dependent rotation angle is the same as that in the single-carrier detector with homodyne readout\,\cite{kimble2001}. Thus one filter cavity is sufficient and its parameters are identical to that in single-carrier detector, as long as $2\omega_m$ is an integer multiple of the free spectral range (FSR) of the filter cavity. In Fig.~\ref{fig:Strain}, we plot the noise spectral densities for different configurations as a comparison. The two-carrier detector with heterodyne readout gives identical sensitivity to that of Advanced LIGO with homodyne readout. The figure also shows that the scheme is robust against a power imbalance between the two carriers. A rather large 15\% power imbalance between the two carriers under 10dB frequency-dependent squeezing only results in a 20\% degradation in sensitivity at low frequencies. At low frequencies, where the radiation pressure noise dominates, an ideal EPR measurement
at the output of the interferometer is not enabled due to asymmetric optomechanical effect from power imbalance. The impact of a 15\% power imbalance on the signal is negligible, leading to only a 0.57\% sensitivity degradation at high frequencies.

\textit{Criteria for macroscopic lengths  --- }
In our proposed scheme, the lengths between core optics need to be carefully set to defined absolute values to guarantee co-resonance of the respective optical fields. This introduces requirements on the macroscopic lengths, in addition to the usual requirements for controlling the microscopic position of the optics. We anticipate this co-resonance requirement to be relatively easy to achieve, as the current lock acquisition system already permits selecting a specific fringe. 

To keep the carriers resonant and the LO beam anti-resonant in the arm cavities, $2\omega_m$ shall be an odd multiple of the FSR of the arm cavity. Another consideration is on the coupling between the symmetric and anti-symmetric ports for both the carriers and the LO. 
Taking the LO field as DC by convention ($\omega_L=0$, $\omega_1=-\omega_m$, $\omega_2=\omega_m$), and locking the central Michelson on its bright fringe, we can treat the central Michelson as an effective mirror with amplitude transmissivity 
\begin{equation}
r_{\rm MI}=i r_a \sin \frac{\omega \Delta l}{c}\,,\quad t_{\rm MI}=r_a  \cos \frac{\omega \Delta l}{c}\,,
\end{equation}
where $\omega$ is the sideband frequency, $\Delta l$ is the Schnupp asymmetry \cite{Izumi_2016}, and $r_a$ is the amplitude reflectivity of arm cavities. For the LO anti-resonating in the arm cavities, $r_a=-1$, and $\omega=\omega_L=0$. For the carriers we have $r_a=1$, as well as $\omega=\omega_1=-\omega_m$ and $\omega=\omega_2=\omega_m$ respectively. To keep the carriers on the Michelson dark fringe we need to have $\omega_m \Delta l/c=\pi/2$.
The macroscopic round-trip length of the signal recycling cavity and power recycling cavity should be tuned to satisfy the following conditions: in the signal recycling cavity, the signal (carrier) modes are anti-resonant, while the LO beam is resonant; in the power recycling cavity, all three beams are on resonance. 
In Advanced LIGO, $\omega_m$ is around $2\pi\times$45\,MHz, so $\Delta l$ needs to be around 1.67\,m. Given the Advanced LIGO power and signal recycling  mirror transmissivities of 0.03 and 0.325, the effective power transmissivity of the LO field from the symmetric port to the anti-symmetric port is around 25\%.
To use one filter cavity for two modes, we can make $2\omega_m$ be an integer multiple of the FSR of the filter cavity, as mentioned earlier. 

\textit{The mirror motion coupling at $2\omega_m$ --- }
With two carriers resonating in the arm cavities, the mirror motion around $2\omega_m$ sensed by one carrier is also measured within the audio-band with respect to other carrier. The additional noise current is
\begin{equation}\begin{split}
I_{2\omega_m}\propto \hat{s}_{1,2\omega_m-\Omega}^{\dagger}e^{i(\phi_L+\phi_D)}+ \hat{s}_{2,-2\omega_m-\Omega}^{\dagger}e^{i(\phi_L-\phi_D)}\\
+\hat{s}_{1,2\omega_m+\Omega}e^{-i(\phi_L+\phi_D)}+ \hat{s}_{2,-2\omega_m+\Omega}e^{-i(\phi_L-\phi_D)}\,,
\end{split}
\end{equation}
where $\phi_L=\frac{\pi}{2}, \phi_D=0$ in our scheme. 
Around frequency $2\omega_m$, the radiation pressure also excites mirror motion at the vibration mode resonances of the mirror. The susceptibility at $2\omega_m+\Omega$ of the vibration mode at resonant frequency $\omega_a$ can be modelled as a harmonic oscillator with damping part described by mechanical dispersion, $\psi$ ($\psi=1/Q$, $Q$ is the quality factor), as\,\cite{PhysRevD.52.577}
\begin{equation}
\chi_a=\frac{1}{\mu m \left[\omega_a^2+i \omega_a^2 \psi -(2\omega_m+\Omega)^2\right]}\,,
\end{equation}
where $\mu$ is an effective mass coefficient which includes the coupling between mirror mode and the carrier Gaussian mode. 

In order to not affect the quantum sensitivity, we need the radiation pressure induced mirror motion around $2\omega_m$ to be much smaller than the quantum shot noise. In other words, $|\chi_a|$ need to be much smaller than the absolute value of the free mass susceptibility, $1/(m\Omega^2)$, at frequency where radiation pressure noise is approximate to shot noise. For $\omega_a=2\omega_m+\Omega$, and taking $\Omega=2\pi\times 60$\,Hz, $L=4$\,km, $\omega_m\approx 2\pi\times$ 45\,MHz, this requires
\begin{equation}\label{eq:Qmu}
\frac{Q}{\mu}\ll 2.25\times 10^{12}\,. 
\end{equation}
Around 90\,MHz, the $Q$ of silica is around $10^5$\,\cite{PhysRevLett.84.2718}. We use program, \textit{Cypres}\,\cite{BONDU199574} and simulate the effective mass coefficients up to 300\,kHz taking Advanced LIGO mirror and beam size as the example. As it turns out $\mu$ is always larger than $0.1$ without observing a trend of smaller $\mu$ towards higher frequency. Details are in Supplementary Material. It proves the satisfaction of Eq.~\ref{eq:Qmu} indirectly. 

The thermal noise at 90\,MHz from the mode at the same frequency can be calculated as\,\cite{PhysRevD.52.577}
\begin{equation}
\begin{split}
\sqrt{S_{th}}&=\sqrt{4\times\frac{4k_{B}TQ}{\mu m \omega_a^3L^2}}\\&=2.4\times10^{-25}\sqrt{\frac{T}{300\, \rm K}\times\frac{1}{\mu}\times\frac{Q}{10^5}} \frac{1}{\sqrt{\rm Hz}}\,,
\end{split}
\end{equation}
with a bandwidth around $\omega_a/(2\pi Q)=900\,\rm Hz$. $k_B$ is the Boltzmann constant, $T$ is the environment temperature. From the mode simulation (with details in Supplementary Material), we observe the density of modes with $\mu$ no larger than 1 is almost constant over frequency and around one mode per 2.5\,kHz in average. Even though assuming that the mode separation is the same as the mode bandwidth, the thermal noise at 90\,MHz results $3.1\times 10^{-25}\frac{1}{\sqrt{\rm Hz}}$, by taking contributions from 100 modes in the vicinity into account.   
The thermal noise is compatible with the quantum noise of the two-carrier detector but experimental studies with more details are required in the future.

\textit{Conclusions and Discussions --- }
We have shown that the proposed two-carrier gravitational wave detector with heterodyne readout evades the 3dB quantum penalty of conventional heterodyne readout. It also allows the usage of two-mode squeezing in the same way of
single mode squeezing with homodyne readout.
Furthermore, the two-carrier detector provides advantages: (1) it can enable squeezing enhanced measurements in the audio-band and below with high-frequency squeezing, which is immune to the audio-band LO scattering contamination to the squeezer, although is still susceptible to the scattering from residual carriers; (2) it allows us to operate the interferometer on the dark fringe without an additional LO path and output mode cleaners that are essential to the balanced homodyne readout scheme, in which two mode cleaners are required~\cite{zhang2020quantum}. If the higher optical modes at the dark port cannot be suppressed by the interferometer itself, one output mode cleaner of which the FSR equals to $\omega_m$ is sufficient.

As an outlook, we want to 
highlight that the two-carrier
detector is compatible with general quantum nondemolition schemes\,\cite{Braginsky547,yanbeiQND}, in contrast to conventional heterodyne readout\,\cite{PhysRevD.67.122005}. For example, similar to the implementation of frequency dependent squeezing, we can add a filter cavity at the output to realise frequency-dependent readout for back action evasion~\cite{kimble2001}.
The resulting optimal sensitivity is 
\begin{equation}
S_{hh}^{\rm opt}=\frac{h_{\rm SQL}^2}{2}\frac{1}{\mathcal{K}}e^{-2r}\,.
\end{equation}
This saturates the fundamental 
quantum limit or the quantum Cram\'er-Rao bound\,\cite{HELSTROM1967101, doi:10.1063/1.1291855,PhysRevLett.106.090401,PhysRevLett.119.050801}. 

{\it Acknowledgements ---} We thank the support from simulation software, \textit{Finesse} and F.B. for providing the simulation program, \textit{Cypres}.
T. Z., P.J, H. M., D. M. and A. F. acknowledge the support of the Institute for Gravitational Wave Astronomy at University of Birmingham. A. F. has been supported by a Royal Society Wolfson Fellowship which is jointly funded by the Royal Society and the Wolfson Foundation. H. M. is supported by UK STFC Ernest Rutherford Fellowship (Grant No. ST/M005844/11). S.W.B. acknowledges the supported by the National Science Foundation award PHY-1912536. This document was assigned the LIGO document control number LIGO-P2000260.

\bibliography{bibliography}
\end{document}